\pdfoutput=1
\documentclass[12pt,a4paper]{article}

\usepackage{ifthen} 
\newboolean{pdflatex}
\setboolean{pdflatex}{true} 

\newboolean{articletitles}
\setboolean{articletitles}{true} 

\newboolean{uprightparticles}
\setboolean{uprightparticles}{false} 

\def\paperauthors{LHCb collaboration} 
\def\paperasciititle{Measurement of forward charged hadron flow harmonics in peripheral PbPb collisions at center-of-mass energy of 5.02 TeV with the LHCb detector} 
\def\papertitle{Measurement of forward charged hadron flow harmonics in peripheral PbPb collisions at $\sqsnn=5.02$\,\tev with the \lhcb detector} 
\def\paperkeywords{{Heavy-ion collisions},{PbPb},{flow},{two-particle correlations}, {LHCb}} 
\def\papercopyright{\the\year\ CERN for the benefit of the LHCb collaboration} 
\def\paperlicence{CC BY 4.0 licence}
\def\paperlicenceurl{https://creativecommons.org/licenses/by/4.0/}

\usepackage[top=1in, bottom=1.25in, left=1in, right=1in]{geometry}

\usepackage{multicol}
\usepackage{multirow}
\usepackage{placeins}
\usepackage{microtype}
\usepackage{lineno}  
\usepackage{xspace} 
\usepackage{caption} 

\usepackage{graphicx}  
\usepackage{color}
\usepackage{colortbl}
\graphicspath{{./figs/}} 

\usepackage{amsmath} 
\usepackage{amssymb}
\usepackage{amsfonts}
\usepackage{upgreek} 

\newcommand*\patchAmsMathEnvironmentForLineno[1]{%
\expandafter\let\csname old#1\expandafter\endcsname\csname #1\endcsname
\expandafter\let\csname oldend#1\expandafter\endcsname\csname
end#1\endcsname
 \renewenvironment{#1}%
   {\linenomath\csname old#1\endcsname}%
   {\csname oldend#1\endcsname\endlinenomath}%
}
\newcommand*\patchBothAmsMathEnvironmentsForLineno[1]{%
  \patchAmsMathEnvironmentForLineno{#1}%
  \patchAmsMathEnvironmentForLineno{#1*}%
}
\AtBeginDocument{%
\patchBothAmsMathEnvironmentsForLineno{equation}%
\patchBothAmsMathEnvironmentsForLineno{align}%
\patchBothAmsMathEnvironmentsForLineno{flalign}%
\patchBothAmsMathEnvironmentsForLineno{alignat}%
\patchBothAmsMathEnvironmentsForLineno{gather}%
\patchBothAmsMathEnvironmentsForLineno{multline}%
\patchBothAmsMathEnvironmentsForLineno{eqnarray}%
}

\usepackage{hyperxmp}

\usepackage[pdftex,
            pdfauthor={\paperauthors},
            pdftitle={\paperasciititle},
            pdfkeywords={\paperkeywords},
            pdfcopyright={Copyright (C) \papercopyright},
            pdflicenseurl={\paperlicenceurl}]{hyperref}

\usepackage[colorinlistoftodos,textsize=scriptsize]{todonotes}

\usepackage[bottom,flushmargin,hang,multiple]{footmisc}

\usepackage[all]{hypcap} 

\usepackage{xspace} 
\usepackage{upgreek}

\def\lhcb   {\mbox{LHCb}\xspace}
\def\atlas  {\mbox{ATLAS}\xspace}

\def\alice  {\mbox{ALICE}\xspace}

\def\lhc    {\mbox{LHC}\xspace}

\def\velo   {VELO\xspace}

\def\MagUp {\mbox{\em Mag\kern -0.05em Up}\xspace}

\ifthenelse{\boolean{uprightparticles}}%
{

 \def\PDelta      {\ensuremath{\Delta}\xspace}                 
 \def\PXi         {\ensuremath{\Xi}\xspace}                 
 \def\PLambda     {\ensuremath{\Lambda}\xspace}                 
 \def\PSigma      {\ensuremath{\Sigma}\xspace}                 
 \def\POmega      {\ensuremath{\Omega}\xspace}                 
 \def\PUpsilon    {\ensuremath{\Upsilon}\xspace}
 \let\oldPi\Pi
 \def\PPi         {\ensuremath{\oldPi}\xspace}

 \def\PB      {\ensuremath{\mathrm{B}}\xspace}                 
                  
 \def\PD      {\ensuremath{\mathrm{D}}\xspace}

 \def\PK      {\ensuremath{\mathrm{K}}\xspace}

 \def\Pb      {\ensuremath{\mathrm{b}}\xspace}                 
 \def\Pc      {\ensuremath{\mathrm{c}}\xspace}

 \def\Pi      {\ensuremath{\mathrm{i}}\xspace}

 \def\Pp      {\ensuremath{\mathrm{p}}\xspace}

 \def\Ps      {\ensuremath{\mathrm{s}}\xspace}

 \def\thebaroffset{0.0em}
}
{

 \mathchardef\PDelta="7101
 \mathchardef\PXi="7104
 \mathchardef\PLambda="7103
 \mathchardef\PSigma="7106
 \mathchardef\POmega="710A
 \mathchardef\PUpsilon="7107
 \mathchardef\PPi="7105
                  
 \def\PB      {\ensuremath{B}\xspace}                 
                  
 \def\PD      {\ensuremath{D}\xspace}

 \def\PK      {\ensuremath{K}\xspace}

 \def\Pb      {\ensuremath{b}\xspace}                 
 \def\Pc      {\ensuremath{c}\xspace}

 \def\Pi      {\ensuremath{i}\xspace}

 \def\Pp      {\ensuremath{p}\xspace}

 \def\Ps      {\ensuremath{s}\xspace}

 \def\thebaroffset{0.18em}
}
\newcommand{\offsetoverline}[2][\thebaroffset]{\kern #1\overline{\kern -#1 #2}}%

\makeatletter
\ifcase \@ptsize \relax
  \newcommand{\miniscule}{\@setfontsize\miniscule{4}{5}}
\or
  \newcommand{\miniscule}{\@setfontsize\miniscule{5}{6}}
\or
  \newcommand{\miniscule}{\@setfontsize\miniscule{5}{6}}
\fi
\makeatother

\DeclareRobustCommand{\optbar}[1]{\shortstack{{\miniscule (\rule[.5ex]{1.25em}{.18mm})}
  \\ [-.7ex] $#1$}}

\def\squark    {{\ensuremath{\Ps}}\xspace}

\def\cquark    {{\ensuremath{\Pc}}\xspace}

\def\bquark    {{\ensuremath{\Pb}}\xspace}

\def\KorKbar {\kern \thebaroffset\optbar{\kern -\thebaroffset \PK}{}\xspace}

\def\D       {{\ensuremath{\PD}}\xspace}

\def\DorDbar {\kern \thebaroffset\optbar{\kern -\thebaroffset \PD}\xspace}

\def\Dp      {{\ensuremath{\D^+}}\xspace}
\def\Dm      {{\ensuremath{\D^-}}\xspace}

\def\DpDm    {\ensuremath{\Dp {\kern -0.16em \Dm}}\xspace}

\def\B       {{\ensuremath{\PB}}\xspace}

\def\BorBbar {\kern \thebaroffset\optbar{\kern -\thebaroffset \PB}\xspace}

\def\Bd      {{\ensuremath{\B^0}}\xspace}

\def\BdorBdbar {\kern \thebaroffset\optbar{\kern -\thebaroffset \Bd}\xspace}

\def\Bs      {{\ensuremath{\B^0_\squark}}\xspace}

\def\BsorBsbar {\kern \thebaroffset\optbar{\kern -\thebaroffset \Bs}\xspace}

\def\Y#1S{\ensuremath{\PUpsilon{(#1S)}}\xspace}

\def\proton      {{\ensuremath{\Pp}}\xspace}

\def\LorLbar     {\kern \thebaroffset\optbar{\kern -\thebaroffset \PLambda}\xspace}

\def\AT#1     {\ensuremath{A_{\mathrm{T}}^{#1}}\xspace}           

\def\C#1      {\ensuremath{\mathcal{C}_{#1}}\xspace}                       
\def\Cp#1     {\ensuremath{\mathcal{C}_{#1}^{'}}\xspace}                    
\def\Ceff#1   {\ensuremath{\mathcal{C}_{#1}^{\mathrm{(eff)}}}\xspace}        
\def\Cpeff#1  {\ensuremath{\mathcal{C}_{#1}^{'\mathrm{(eff)}}}\xspace}       
\def\Ope#1    {\ensuremath{\mathcal{O}_{#1}}\xspace}                       
\def\Opep#1   {\ensuremath{\mathcal{O}_{#1}^{'}}\xspace}                    

\newcommand{\nospaceunit}[1]{\ensuremath{\text{#1}}}       
\newcommand{\aunit}[1]{\ensuremath{\text{\,#1}}}       

\newcommand{\tev}{\aunit{Te\kern -0.1em V}\xspace}
\newcommand{\gev}{\aunit{Ge\kern -0.1em V}\xspace}
\newcommand{\mev}{\aunit{Me\kern -0.1em V}\xspace}
\newcommand{\kev}{\aunit{ke\kern -0.1em V}\xspace}
\newcommand{\ev}{\aunit{e\kern -0.1em V}\xspace}
 
\newcommand{\mevc}{\ensuremath{\aunit{Me\kern -0.1em V\!/}c}\xspace}
\newcommand{\gevc}{\ensuremath{\aunit{Ge\kern -0.1em V\!/}c}\xspace}
\newcommand{\mevcc}{\ensuremath{\aunit{Me\kern -0.1em V\!/}c^2}\xspace}
\newcommand{\gevcc}{\ensuremath{\aunit{Ge\kern -0.1em V\!/}c^2}\xspace}

\def\mum  {\ensuremath{\,\upmu\nospaceunit{m}}\xspace}

\def\invmub  {\ensuremath{\,\upmu\nospaceunit{b}^{-1}}\xspace}

\def\gsim{{~\raise.15em\hbox{$>$}\kern-.85em
          \lower.35em\hbox{$\sim$}~}\xspace}
\def\lsim{{~\raise.15em\hbox{$<$}\kern-.85em
          \lower.35em\hbox{$\sim$}~}\xspace}

\def\sqsnn {\ensuremath{\protect\sqrt{s_{\scriptscriptstyle\text{NN}}}}\xspace}
\def\pt         {\ensuremath{p_{\mathrm{T}}}\xspace}

\def\ptot       {\ensuremath{p}\xspace}

\def\evtgen     {\mbox{\textsc{EvtGen}}\xspace}

\def\geant      {\mbox{\textsc{Geant4}}\xspace}

\def\photos     {\mbox{\textsc{Photos}}\xspace}

\def\tell1  {TELL1\xspace}
\def\ukl1   {UKL1\xspace}

\newcommand{\lhcborcid}[1]{\href{https://orcid.org/#1}{\hspace*{0.1em}\raisebox{-0.45ex}{\includegraphics[width=1em]{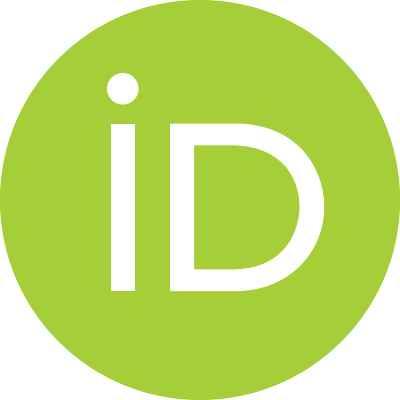}}}}

\usepackage{cite} 
\usepackage{mciteplus}

\usepackage{longtable} 

\usepackage{enumitem}
\usepackage{url}

\begin{document}

\renewcommand{\thefootnote}{\fnsymbol{footnote}}
\setcounter{footnote}{1}


\begin{titlepage}
\pagenumbering{roman}

\vspace*{-1.5cm}
\centerline{\large EUROPEAN ORGANIZATION FOR NUCLEAR RESEARCH (CERN)}
\vspace*{1.5cm}
\noindent
\begin{tabular*}{\linewidth}{lc@{\extracolsep{\fill}}r@{\extracolsep{0pt}}}
\ifthenelse{\boolean{pdflatex}}
{\vspace*{-1.5cm}\mbox{\!\!\!\includegraphics[width=.14\textwidth]{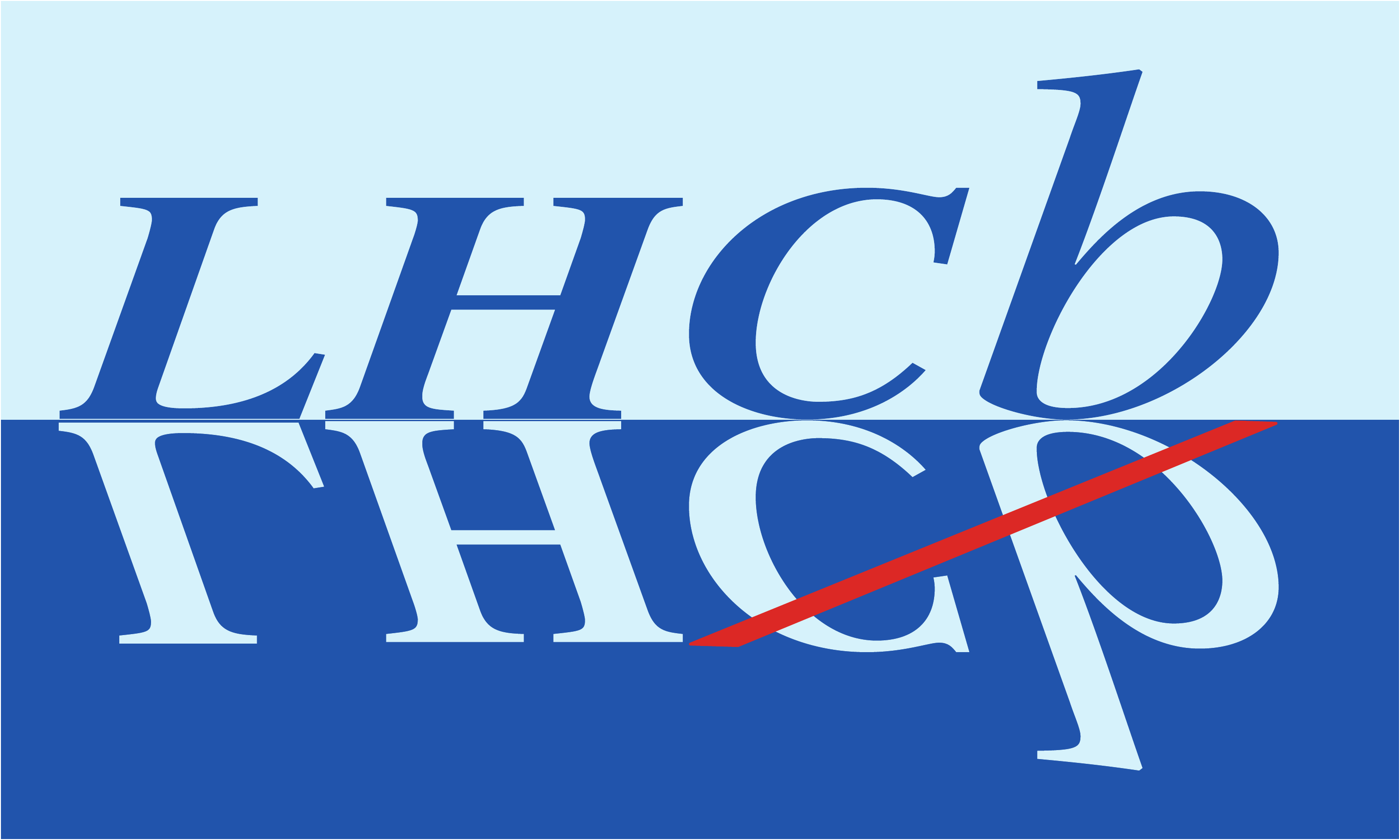}} & &}%
{\vspace*{-1.2cm}\mbox{\!\!\!\includegraphics[width=.12\textwidth]{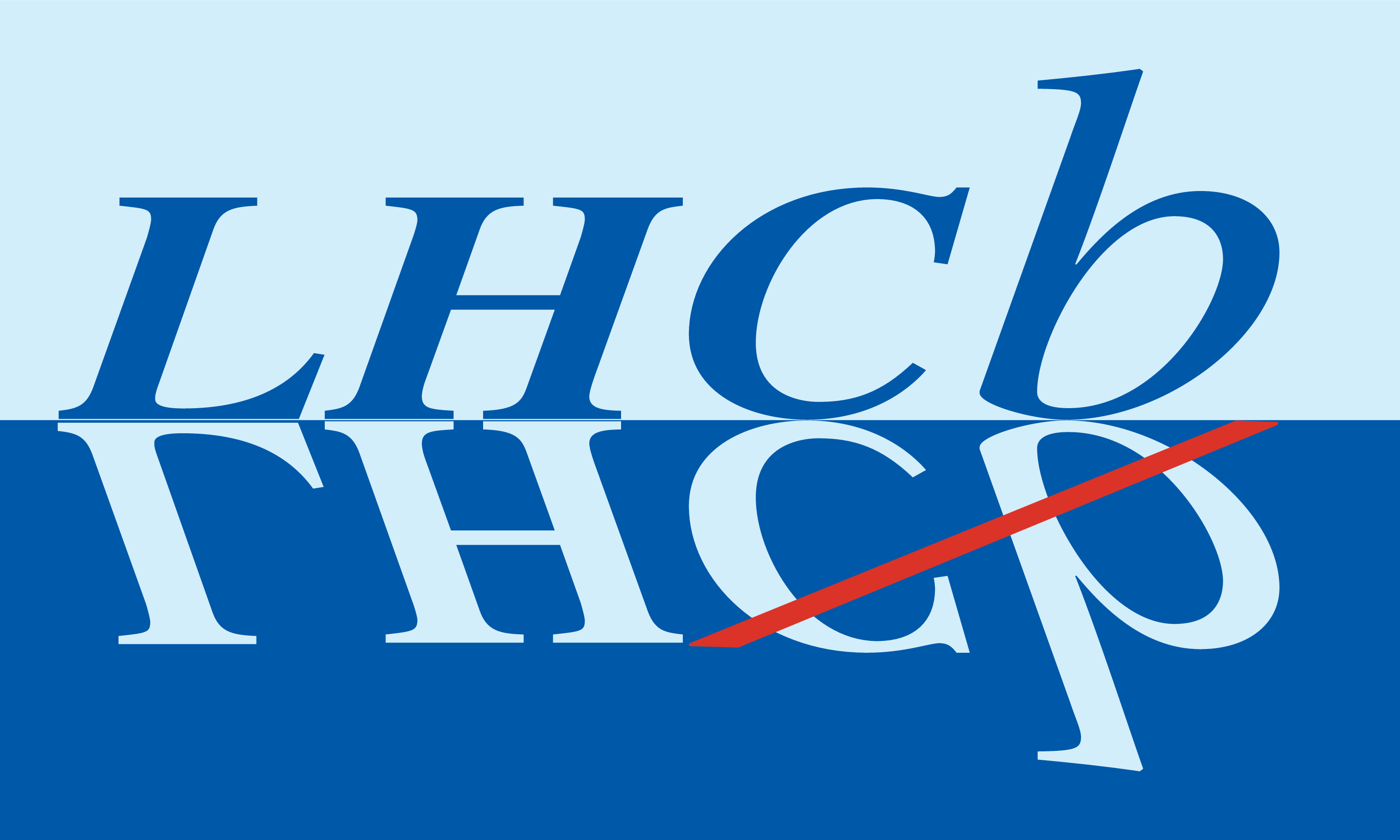}} & &}%
\\
 & & CERN-EP-2023-240 \\  
 & & LHCb-PAPER-2023-031 \\  
 & & May 15, 2024 \\
 & & \\
\end{tabular*}

\vspace*{2.0cm}

{\normalfont\bfseries\boldmath\huge
\begin{center}
  \papertitle 
\end{center}
}

\vspace*{2.0cm}

\begin{center}
\paperauthors\footnote{Authors are listed at the end of this paper.}
\end{center}

\vspace{\fill}

\begin{abstract}
  Flow harmonic coefficients, $v_n$, which are the key to studying the hydrodynamics of the quark-gluon plasma (QGP) created in heavy-ion collisions, have been measured in various collision systems and kinematic regions and using various particle species. The study of flow harmonics in a wide pseudorapidity range is particularly valuable to understand the temperature dependence of the shear viscosity to entropy density ratio of the QGP. This paper presents the first LHCb results of the second- and the third-order flow harmonic coefficients of charged hadrons as a function of transverse momentum in the forward region, corresponding to pseudorapidities between 2.0 and 4.9, using the data collected from PbPb collisions in 2018 at a center-of-mass energy of $5.02$\,\tev. The coefficients measured using the two-particle angular correlation analysis method are smaller than the central-pseudorapidity measurements at \alice and \atlas from the same collision system but share similar features.
  
\end{abstract}

\vspace*{2.0cm}

\begin{center}
  Phys.~Rev.~C 109 (2024), 054908
\end{center}

\vspace{\fill}

{\footnotesize 
\centerline{\copyright~\papercopyright. \href{\paperlicenceurl}{\paperlicence}.}}
\vspace*{2mm}

\end{titlepage}


\newpage
\setcounter{page}{2}
\mbox{~}
%
%
%
%


\renewcommand{\thefootnote}{\arabic{footnote}}
\setcounter{footnote}{0}

\cleardoublepage


\pagestyle{plain} 
\setcounter{page}{1}
\pagenumbering{arabic}


\section{Introduction}
Quark-gluon plasma (QGP) is a phase of nuclear matter in which partons can move freely, as explained by the asymptotic freedom of quantum chromodynamics (QCD). The QGP medium is formed in an extremely hot and dense environment, such as in high energy collisions of heavy-ions~\cite{ADCOX2005184,KALASHNIKOV1979328,STAR:2005gfr,PHOBOS:2004zne,BRAHMS:2004adc}. As the heavy ions collide at near the speed of light, a dense QGP medium forms and thermalizes rapidly.

The unbound partons of the QGP move collectively. This collective movement is heavily affected by the initial collision conditions, such as the momentum anisotropy due to the asymmetric collision geometry. These conditions cause spatial anisotropy in the final particle distributions. The study of the spatial anisotropy, commonly known as flow, helps us to understand the evolution and the properties of the QGP, including the thermalization process, initial- and final-state effects, and the transport properties including the ratio of shear viscosity to entropy density. The value of the ratio of shear viscosity to entropy density is found to be small~\cite{physRevLett.94.111601}, which indicates that the QGP medium behaves like a nearly perfect fluid.

Several theory models predict that the ratio of shear viscosity to entropy density is temperature dependent~\cite{etaOvers_LHC_RHIC,PhysRevC.90.044904,PhysRevLett.116.212301}. The study of this temperature dependence requires a wide range of particle flow measurements, covering different collision energies, centralities, transverse momenta and pseudorapidities, such as those reported by experiments at both the BNL Relativistic Heavy Ion Collider (RHIC) and the CERN Large Hadron Collider (LHC). Measurements to date include \proton\proton, $d$Au, $^3$HeAu, XeXe, AuAu and PbPb collision systems~\cite{PHENIX_Nature,STAR_smallLargeSys,CMS_pPb_PbPb,ATLAS_XeXe,ALICE_PbPb,ALICE_idenifiedHadronV2}, energies ranging from $\mathcal{O}(\gev)$ to $\mathcal{O}(\tev)$~\cite{STAR_energy,STAR_energy2,ALICE_vnCent}, and the most central events to ultraperipheral collisions~\cite{ALICE_vnCent,ATLAS_UPC}. 
Most of these results are presented in the central pseudorapidity region, $|\eta|<2.5$. Several forward (large~$|\eta|$) particle flow studies were reported by PHOBOS using AuAu collisions at the \gev energy scale~\cite{PHOBOS:2002vby,PHOBOS_v1} and by ALICE using PbPb and \proton{Pb} collisions at the \tev energy scale~\cite{ALICEv2v3,ALICE_jpsiV2,ALICE_Jpsiv2_PbPb}. 

The LHCb experiment can provide unique flow measurements in the forward region, which are important to understand the ``cooler" region where freeze-out is dominant~\cite{PhysRevC.90.044904}. The forward region is dominated by the nonequilibrium hadronic phase and can test the limit of the hydrodynamic and the transport models that describe QGP at microscopic and macroscopic scales, respectively. \lhcb measurements also complement other \lhc results that are in the central-pseudorapidity regions in the effort to constrain theoretical models and understand the evolution of QGP.

This paper reports the first measurement of the forward flow harmonic coefficient of charged hadrons as a function of transverse momentum at \lhcb and at the \lhc to enrich the study of flow in the nonequilibrium hadronic phase of the system evolution. The two-dimensional [$C(\Delta\eta,\Delta\phi)$] and one-dimensional [$C(\Delta\phi)$] correlation functions are constructed using a two-particle correlation analysis method~\cite{CMS_pPb_PbPb,ATLAS_v2pt,ALICE_2PC}. The one-dimensional azimuthal correlation functions are described by a Fourier series, which is used to extract the second- and third-order flow harmonic coefficients, $v_{2}$ and $v_{3}$. The two-particle correlation analysis method is applied to data in two centrality ranges, $65$\%--$75$\% and $75$\%--$84$\%, determined in Ref.~\cite{LHCb-DP-2021-002}. The flow harmonic coefficients are measured as a function of transverse momentum, \pt. The results are compared to those from the \alice and \atlas experiments at central pseudorapidity in PbPb collisions at $5.02$\,\tev~\cite{ALICE_vnCent, ATLAS_v2pt} and to simulation based on the multiphase transport model AMPT~\cite{AMPT1,AMPT2}.

\section{The LHC{b} Experiment}
The analysis is based on data collected with the LHCb detector during the lead-lead data-taking period in 2018. The LHC provided PbPb collisions at a nucleon--nucleon center-of-mass energy of $\sqrt{s_{NN}}=5.02$\,\tev, corresponding to an integrated luminosity of $214$\invmub.

The \lhcb detector~\cite{LHCb-DP-2008-001,LHCb-DP-2014-002} is a single-arm forward spectrometer covering the \mbox{pseudorapidity} range $2<\eta <5$, designed for the study of particles containing \bquark or \cquark quarks. The detector includes a high-precision tracking system consisting of a silicon-strip vertex detector (VELO), surrounding the PbPb interaction region~\cite{LHCb-DP-2014-001}, a large-area silicon-strip detector located upstream of a dipole magnet with a bending power of about $4{\mathrm{\,Tm}}$, and three stations of silicon-strip detectors and straw drift tubes~\cite{LHCb-DP-2013-003,LHCb-DP-2017-001} placed downstream of the magnet. The tracking system provides a measurement of the momentum, \ptot, of charged particles with a relative uncertainty that varies from 0.5\% at low momentum to 1.0\% at 200\gevc. The minimum distance of a track to a primary PbPb collision vertex (PV), the impact parameter (IP), is measured with a resolution of $(15+29/\pt)\mum$, with \pt in\,\gevc. Photons, electrons and hadrons are identified by a calorimeter system consisting of scintillating-pad and preshower detectors (SPD), an electromagnetic calorimeter and a hadronic calorimeter. The online event selection is performed by a trigger~\cite{LHCb-DP-2012-004}, which consists of a hardware stage, based on information from the calorimeter and muon systems, followed by a software stage, which applies a full event reconstruction.

Simulation samples are required to model the effects of the detector acceptance and the selection requirements. The PbPb events are generated using the EPOS event generator~\cite{EPOS} and calibrated with LHC data~\cite{LHCbSim}. Decays of unstable particles are described by \evtgen~\cite{Lange:2001uf}, in which final-state radiation is generated using \photos~\cite{davidson2015photos}. The interaction of the generated particles with the detector, and its response, are implemented using the \geant toolkit~\cite{Allison:2006ve, *Agostinelli:2002hh} as described in Ref.~\cite{LHCb-PROC-2011-006}.

\section{Data selection}\label{sec:dataSelection}
The PbPb events must satisfy at least one of four minimum-bias triggers, all of which place requirements on the number of SPD hits. Two of them impose further requirements based on information from either the hadronic calorimeter or the muon system. Due to hardware limitations, only events with fewer than ten thousand VELO clusters are recorded, which corresponds to centrality greater than $60$\%.

Events with a PV within $\pm3\,\sigma$ of the mean PV $z$ coordinate, where $\sigma$ is the width of the PV $z$ distribution for the dataset, and a centrality between $65$--$84$\% determined by the total calorimeter energy~\cite{LHCb-DP-2021-002} are selected. Centrality ranges below $65$\% are avoided due to lack of sufficient events. The upper bound of the centrality selection is set to avoid contamination with ultraperipheral events~\cite{LHCb-DP-2021-002}. The data are contaminated with fixed-target PbNe collisions that were running simultaneously with the PbPb collisions. Since the PbNe events have lower center-of-mass energy and lower average multiplicity, the selected PbPb events  must have at least $15$~tracks in the backward ($\eta<-2$) region and are required to have a minimum total calorimeter energy that depends on the number of \velo clusters.

All tracks selected in this analysis are measured by the \velo, the silicon-strip detectors located upstream and downstream of the magnet, and the straw drift tubes. These tracks have a minimum momentum of $2$\gevc. The selected tracks must have $\pt > 0.2$\gevc, $2 < \eta < 4.9$, and a small fit $\chi^2$. Tracks from the decays of heavy-flavor hadrons are suppressed by a requirement on the change in the primary vertex $\chi^2$ when the track is excluded from the vertex fit.

\section{Two-particle angular correlation analysis}
Two-particle correlation analysis is based on the fact that the correlations among the produced particles reflect the correlations between the produced particles and the reaction plane~\cite{flowAna}, that is the azimuth of the impact parameter.  Two tracks from the same event, labeled $a$ and $b$, are paired to construct the two-dimensional angular distributions, $S(\Delta\eta,\Delta\phi)$, where $\Delta\eta=\eta_a-\eta_b$ and  $\Delta\phi=\phi_a-\phi_b$. The transverse momentum of track~$a$ is in one of several \pt ranges defined within $0.2 < {\pt}_a < 10$\gevc, but that of track~$b$ must be within $0.2 < {\pt}_b < 5$\gevc regardless of ${\pt}_a$. The lower bound of ${\pt}_b$ is set according to the tracking performance of the detector~\cite{LHCb-PAPER-2021-015}, while the upper bound is set to reduce jet-like contributions at high \pt~\cite{ATLAS_v2pt}. Applying the same track requirements, two tracks from different events are also paired to construct the mixed-event angular distributions, $B(\Delta\eta,\Delta\phi)$, which carry any biases from the detector acceptance.

The two-dimensional angular correlation functions, $C(\Delta\eta,\Delta\phi)$, are obtained by correcting the same-event correlations using the mixed-event correlations, such~that

\begin{align}
    C(\Delta\eta,\Delta\phi)&=\frac{S(\Delta\eta,\Delta\phi)}{B(\Delta\eta,\Delta\phi)}\text{ .}
\end{align}

\begin{figure}[ht]
    \centering
    \includegraphics[width=0.8\textwidth]{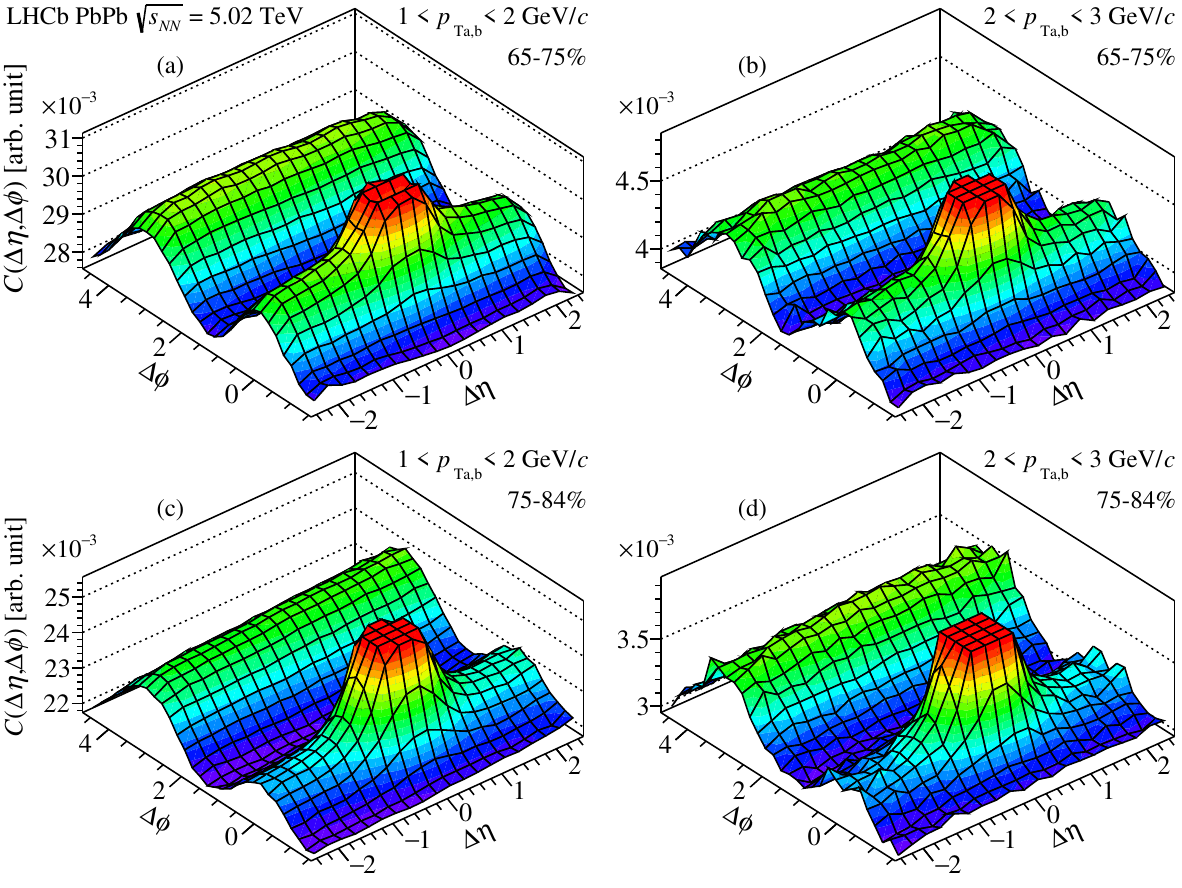}
    \caption{Angular correlation functions in four example intervals of transverse momentum and centrality. The $\Delta\eta$ range is limited to $\pm2.5$. The $z$ axis is cropped to visualize the ridge structures.}
    \label{fig:corr2D}
\end{figure}

Figure~\ref{fig:corr2D} shows an example of the two-dimensional angular correlation functions for ${1<{\pt}_{a,b}<2}$\gevc and $2<{\pt}_{a,b}<3$\gevc and in centrality ranges $65$\%--$75$\% and $75$\%--$84$\%. These two ${\pt}_{a,b}$ ranges are selected to match Ref.~\cite{LHCB-PAPER-2015-040} for comparisons, but the rest of the analysis shown below used the fixed ${\pt}_{b}$ range of 0.2--5\gevc. A clear near-side peak at $(\Delta\eta,\Delta\phi)=(0,0)$, which arises from the short-range nonflow contributions~\cite{ATLAS_factBreaking}, such as jets, is observed in both $\pt$ ranges and centrality ranges. A dip at the center of the near-side peak is observed in the low $\pt$ range in the $65$--$75$\% centrality range. This dip is caused by the removal of tracks that share $70$\% of hits in the track reconstruction~\cite{LHCbForwardTrking}. There are ridge structures on the near ($\Delta\phi\approx0$) and away side ($\Delta\phi\approx\pi$). The near-side ridges, which are less noticeable than the away-side ridges, are a sign of particle flow~\cite{LHCB-PAPER-2015-040}. The near-side ridges are more pronounced in Fig.~\ref{fig:corr2D} compared to the correlation functions in \proton{Pb} and Pb\proton collisions in Ref.~\cite{LHCB-PAPER-2015-040}, indicating stronger flow in PbPb collisions.

One-dimensional azimuthal correlation functions, $C(\Delta\phi)$, are obtained by taking the ratio of the projection of $S(\Delta\eta,\Delta\phi)$ and $B(\Delta\eta,\Delta\phi)$ onto the $\Delta\phi$ axis. The $|\Delta\eta|<1$ region is removed in the projection to reduce short-range nonflow contributions, such that the azimuthal correlation function is 

\begin{align}
    C(\Delta\phi)=\frac{\int^{2.9}_1 S(|\Delta\eta|,\Delta\phi)\cdot d(|\Delta\eta|)}{\int^{2.9}_1 B(|\Delta\eta|,\Delta\phi)\cdot d(|\Delta\eta|)}\text{ .}
\end{align}

A Fourier series fit to this function is performed including the first three harmonic terms,

\begin{align}
    C(\Delta\phi)=A\left[1+2\sum^{3}_{n=1} V_n(\langle{\pt}_a\rangle,\langle{\pt}_b\rangle)\cos{(n\cdot\Delta\phi)}\right]\text{ ,}\label{eq:fit}
\end{align}
where $\langle{\pt}_a\rangle$ ($\langle{\pt}_b\rangle$) represents the average ${\pt}_a$ (${\pt}_b$) for the given ${\pt}_a$ (${\pt}_b$) range, $A$ and $V_n(\langle{\pt}_a\rangle,\langle{\pt}_b\rangle)$ are parameters that vary freely in the fit. The coefficient  $V_n(\langle{\pt}_a\rangle,\langle{\pt}_b\rangle)$ extracted from the fit can be factorized as 

\begin{align}
V_n(\langle{\pt}_a\rangle,\langle{\pt}_b\rangle)=v^a_n(\langle{\pt}_a\rangle)\cdot v^b_n(\langle{\pt}_b\rangle)\text{ ,}\label{eq:fact}
\end{align}
where $v^a_n(\langle{\pt}_a\rangle)$ ($v^b_n(\langle{\pt}_b\rangle)$) is the $n$\textsuperscript{th} flow harmonic coefficient of particle~$a$~($b$) with a transverse momentum in $\langle{\pt}_a\rangle$ ($\langle{\pt}_b\rangle$)~\cite{ALICE_factBreaking}. Since the ${\pt}_b$ range is fixed regardless of the  ${\pt}_a$ interval, one can first obtain $v^b_n(\langle{\pt}_b\rangle)$ by constructing the azimuthal correlations in Eq.~\eqref{eq:fit} of tracks from the $b$ tracks only. Then, Eq.~\eqref{eq:fact} becomes

\begin{align}
V_n(\langle{\pt}_b\rangle,\langle{\pt}_b\rangle)&=v^b_n(\langle{\pt}_b\rangle)\cdot v^b_n(\langle{\pt}_b\rangle)\text{ ,}
\end{align}
which implies
\begin{align}
v^b_n(\langle{\pt}_b\rangle)&=\sqrt{V_n(\langle{\pt}_b\rangle,\langle{\pt}_b\rangle)}\text{ .}\label{eq:vnb}
\end{align}
The flow harmonic coefficient of particle~$a$, $v^a_n(\langle{\pt}_a\rangle)$, from the track $a$-$b$ azimuthal correlations is obtained by substituting Eq.~\eqref{eq:vnb} into Eq.~\eqref{eq:fact}.

However, this factorization in Eqs.~\eqref{eq:fact} and~\eqref{eq:vnb} does not apply to the first-order flow harmonic coefficient, as it is strongly affected by the long-range nonflow contributions~\cite{ATLAS_factBreaking,ALICE_factBreaking}. Therefore, the first-order flow harmonic coefficient, $v_1$, is not reported. These long-range nonflow contributions may also affect the higher-order flow harmonic coefficients in peripheral events at high \pt~\cite{ATLAS_v2pt,ATLAS_factBreaking}. These effects include the increase and decrease of the even- and odd-order flow harmonic coefficients, respectively, at high \pt. Since only $v^a_n(\langle{\pt}_a\rangle)$ results are shown, in the rest of this paper $v_n$ and \pt denote $v^a_n$ and $\langle{\pt}_a\rangle$, respectively.

\begin{figure}[ht]
    \centering
    \includegraphics[width=\textwidth]{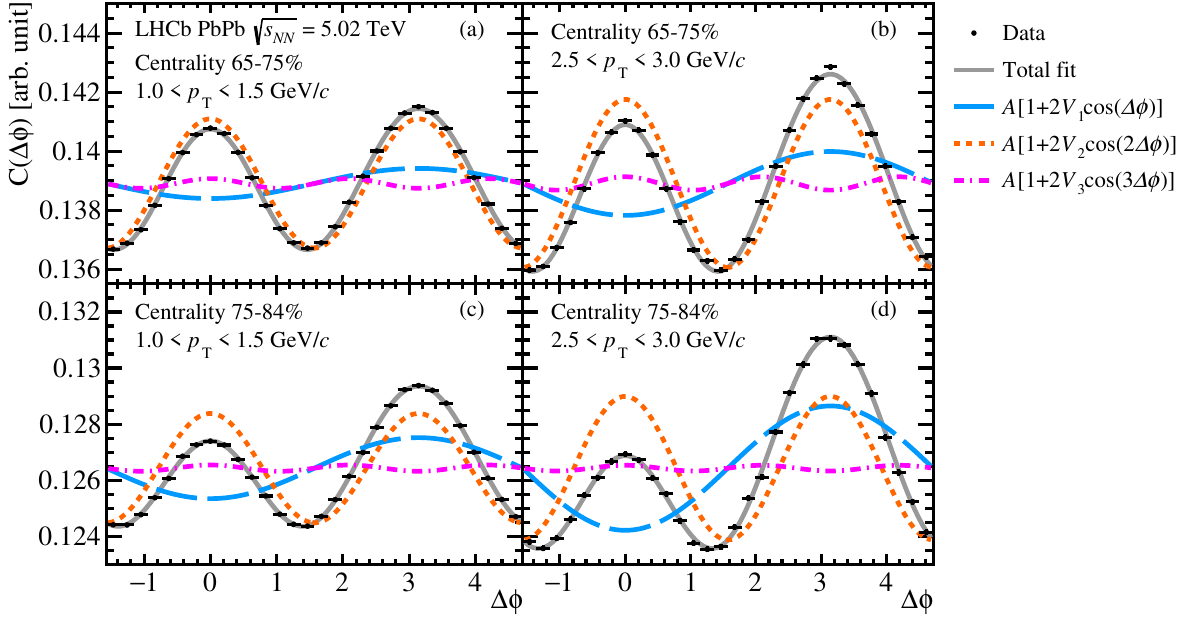}
    \caption{Azimuthal correlation functions in different transverse momentum and centrality ranges. The Fourier series fit and its three terms are overlaid. Only statistical uncertainties are shown.}
    \label{fig:corr}
\end{figure}

Figure~\ref{fig:corr} shows examples of the azimuthal correlation functions overlaid with the Fourier series fit results in different ${\pt}$ and centrality ranges. The relative difference in amplitudes between the near- and away-side peaks is enhanced at high \pt and in peripheral events. The second- and third-order flow harmonic coefficients, $v_2({\pt})$ and $v_3({\pt})$, are extracted from the Fourier series fits in different \pt and centrality ranges.

\section{Systematic uncertainties}
The total systematic uncertainty is obtained from the sum in quadrature of the following six contributions: (a) the primary vertex requirement, (b) the track fit quality requirement, (c) the total calorimeter energy versus \velo multiplicity requirements for PbNe event contamination, (d) Fourier fit fluctuation, (e) fluctuation of the mixed-event correlations, and (f) the unidentified charged hadron efficiency and fake track rate. These systematic uncertainties are estimated by taking the difference of the nominal result and results obtained with alternate requirements as follows.

\begin{enumerate}[label=(\alph*)]
    \item The primary vertex $z$ requirement is varied between $2$ and $4$ standard deviations of the width of the distribution.
    \item The track fit $\chi^2$ requirement is tightened and relaxed such that the \pt distribution with the tightened or relaxed requirement is on average $10$\% different compared to the default requirement.
    \item Besides the minimum total calorimeter energy requirement in the event selection, an additional maximum total calorimeter energy requirement that depends on the number of \velo clusters is applied to remove outlier events with high total calorimeter energy but low multiplicity.
    \item A fourth-order harmonic term is added to the Fourier series fit in Eq.~\eqref{eq:fit}.
    \item The analysis is repeated by moving each data point individually in the mixed-event correlations to its upper and lower statistical limits.
    \item The nominal values of the $v_n$ measurements are obtained without correcting for the detector efficiency, $\epsilon(\pt,\eta)$, and the fake track rate, $f(\pt,\eta)$. To estimate the systematic uncertainties due to the detector efficiency and the fake track rate, the analysis is repeated with the detector efficiency and the fake track rate corrections as a track-paired weight, $w$. The track-paired weight, which is applied when filling two-dimensional angular distributions, is written as

    \begin{align}
       w&=\frac{1-f({\pt}_a,{\eta}_a)}{\epsilon({\pt}_a,{\eta}_a)}\times\frac{1-f({\pt}_b,{\eta}_b)}{\epsilon({\pt}_b,{\eta}_b)}\text{.}
    \end{align}
\end{enumerate}

The relative systematic uncertainties due to the above sources are summarized in Table~\ref{tab:sys} for three \pt ranges:\footnote{The final results are presented with $12$ \pt bins.} $0.2<\pt<0.4$\gevc, $0.4<\pt<3$\gevc, and $3<\pt<10$\gevc. The relative uncertainties are larger in $0.2<\pt<0.4$\gevc and in $3<\pt<10$\gevc, where the nominal $v_n$ is closer to zero or the statistics is low. Furthermore, the relative uncertainties of $v_3$ are generally larger than those of $v_2$ since $v_3$ is closer to zero than $v_2$. Uncertainty sources (b) track fit quality and (f) hadron efficiency and fake track rate are two major contributors to the systematic uncertainties of $v_2$ and $v_3$ in all three \pt ranges. Uncertainty sources (a) primary vertex requirement and (e) fluctuation of mixed-event correlations are subdominant for $v_3$ in the ranges $0.2<\pt<0.4$\gevc and $3<\pt<10$\gevc.

\begin{table}[htbp]
\begin{center}
\caption{Summary of relative systematic uncertainties rounded to the closest 1\%.}
\label{tab:sys}
\begin{tabular}{l | c c c c c c}
\hline
\multicolumn{7}{c}{$v_2$ in $65$--$75$\%} \\ \hline
$p_T$ (\gevc) & $\sigma_a$ & $\sigma_b$ & $\sigma_c$ & $\sigma_d$ & $\sigma_e$ & $\sigma_f$ \\
0.2--0.4 & 4\% & 19--22\% & $<1$\% & $<1$\% & $<1$\% & 6\%\\
0.4--3 & $<1$\% & $<5$\% & $<1$\% & $<1$\% & $<1$\% & $<5$\%\\
3--10 & $<4$\% & $<5$\% & $<1$\% & $<1$\% & $<1$\% & $<10$\%\\
\hline
\multicolumn{7}{c}{$v_2$ in $75$--$84$\%} \\ \hline
$p_T$ (\gevc) & $\sigma_a$ & $\sigma_b$ & $\sigma_c$ & $\sigma_d$ & $\sigma_e$ & $\sigma_f$ \\
0.2--0.4 & $<1$\% & 17--18\% & $<1$\% & $<1$\% & $<1$\% & 27\%\\
0.4--3 & $<1$\% & $<3$\% & $<1$\% & $<1$\% & $<1$\% & $<6$\%\\
3--10 & $<1$\% & $<4$\% & $<1$\% & $<1$\% & $<1$\% & 1--17\%\\
\hline
\multicolumn{7}{c}{$v_3$ in $65$--$75$\%} \\ \hline
$p_T$ (\gevc) & $\sigma_a$ & $\sigma_b$ & $\sigma_c$ & $\sigma_d$ & $\sigma_e$ & $\sigma_f$ \\
0.2--0.4 & 14\% & 23--34\% & 2\% & $<1$\% & 5\% & 64\%\\
0.4--3 & $<5$\% & $<19$\% & $<2$\% & $<1$\% & $<2$\% & 3--9\%\\
3--10 & 2--54\% & $<152$\% & $<9$\% & $<2$\% & 4--22\% & 7--133\%\\
\hline
\multicolumn{7}{c}{$v_3$ in $75$--$84$\%} \\ \hline
$p_T$ (\gevc) & $\sigma_a$ & $\sigma_b$ & $\sigma_c$ & $\sigma_d$ & $\sigma_e$ & $\sigma_f$ \\
0.2--0.4 & 13\% & 21--28\% & 1\% & $<1$\% & 2\% & 14\%\\
0.4--3 & $<9$\% & $<20$\% & $<2$\% & $<1$\% & 1--4\% & 2--35\%\\
3--10 & 1--56\% & 21--142\% & 1--20\% & 1--8\% & 5--31\% & 18--146\%\\
\end{tabular}
\end{center}
\end{table}

%
\section{Results}
\begin{figure}[ht]
    \centering
    \includegraphics[width=0.7\textwidth]{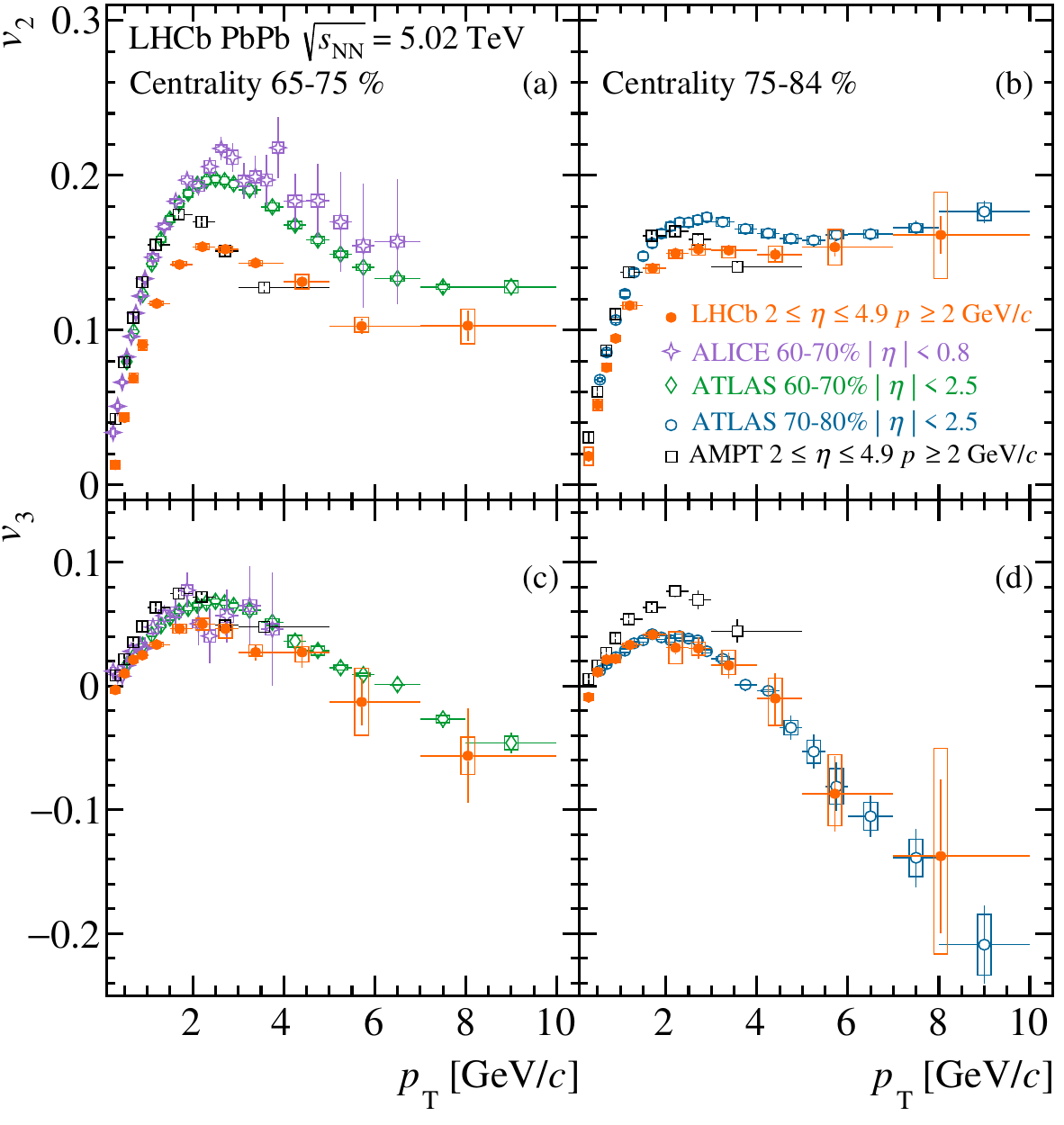}
    \caption{Second- and third-order flow harmonic coefficients as functions of transverse momentum. The statistical and systematic uncertainties are drawn as error bars and boxes, respectively. Only statistical uncertainties are shown for the AMPT predictions.}
    \label{fig:vnPt}
\end{figure}

Figure~\ref{fig:vnPt} shows the measured second- and third-order forward flow harmonic coefficients, $v_2$ and $v_3$, as a function of \pt in PbPb collisions at center-of-mass energy of $5.02$\,\tev. The numerical results are given in the Appendix. These data are compared to ALICE and ATLAS results~\cite{ALICE_vnCent,ATLAS_v2pt} and AMPT simulations~\cite{AMPT1,AMPT2}. The second- and third-order flow harmonic coefficients rise at low \pt and then turn downward after $2.5$\gevc. Above $5$\gevc, the $v_2$ values are consistent as the uncertainties increase at high \pt. Unlike $v_2$, $v_3$ continues to decrease at \pt greater than $2.5$\gevc and goes below zero at \pt greater than $5$\gevc. The consistent $v_2$ and continuous falling of $v_3$ at high \pt hint at factorization breaking due to residual nonflow contributions at high \pt~\cite{ATLAS_v2pt,ATLAS_factBreaking}.

The \alice and \atlas results for $v_2$ and $v_3$ at central pseudorapidity are also extracted from PbPb collision data at center-of-mass energy of $5.02$\,\tev in centrality ranges of $60$\%--$70$\% and $70$\%--$80$\%. The \alice and \atlas results are obtained using the two-particle cumulants and two-particle correlation analysis methods, respectively. These results share similar features, but higher values compared to this paper due to differences in pseudorapidity ranges. This pseudorapidity dependence has also been observed by the PHOBOS~\cite{PHOBOS:2002vby} and \alice~\cite{ALICEv2v3} experiments. 

AMPT simulates particle flow with a string melting model that produces a dense system of partonic matter, and includes quark coalescence to improve the modeling of elliptic flow~\cite{AMPT2}. The AMPT simulations with 68.5-million events overestimate the forward $v_2$ at ${\pt<2.5}$\gevc, and the forward $v_3$ at ${\pt<5}$\gevc. These LHCb data can be used to tune the AMPT~model.

\section{Summary}
This paper presents the first measurements of flow harmonic coefficients of charged hadrons as a function of transverse momentum in the forward direction using PbPb collision data at center-of-mass energy of $5.02$\,\tev. A two-particle angular correlation analysis is used to construct the two-dimensional, $C(\Delta\eta,\Delta\phi)$, and one-dimensional, $C(\Delta\phi)$, correlation functions. The two-dimensional correlations in PbPb show pronounced near- and away-side ridges compared to the published \lhcb~\proton{Pb} and Pb\proton results~\cite{LHCB-PAPER-2015-040}, indicating stronger forward particle flow in PbPb events than in \proton{Pb} and Pb\proton events. The one-dimensional azimuthal correlation functions are used to extract the second- and third-order harmonic coefficients, $v_2$ and $v_3$, as a function of \pt in various centrality ranges.

These $v_2$ and $v_3$ values are generally smaller than those measured by the \alice and the \atlas experiments at central pseudorapidity, which could be due to the dominant freeze-out phase in the forward region leading to weaker flow. However, both studies share the same features of rising $v_2$ and $v_3$ at $\pt<2.5$\gevc, and falling at high \pt. The consistent $v_2$ and the continuous falling of $v_3$ at high \pt may be caused by nonflow contributions or limited statistics at high \pt. The AMPT simulations overestimate both $v_2$ and $v_3$, suggesting that they require tuning.

These $v_2$ and $v_3$ results in the forward direction from LHCb at the \tev energy scale along with other flow measurements at central pseudorapidity will help constrain the theory models of particle flow, and understand the evolution of QGP from the partonic phase to the hadronic phase.

\section*{Acknowledgements}
%
%
\noindent We express our gratitude to our colleagues in the CERN
accelerator departments for the excellent performance of the LHC. We
thank the technical and administrative staff at the LHCb
institutes.
We acknowledge support from CERN and from the national agencies:
CAPES, CNPq, FAPERJ and FINEP (Brazil); 
MOST and NSFC (China); 
CNRS/IN2P3 (France); 
BMBF, DFG and MPG (Germany); 
INFN (Italy); 
NWO (Netherlands); 
MNiSW and NCN (Poland); 
MCID/IFA (Romania); 
MICINN (Spain); 
SNSF and SER (Switzerland); 
NASU (Ukraine); 
STFC (United Kingdom); 
DOE NP and NSF (USA).
We acknowledge the computing resources that are provided by CERN, IN2P3
(France), KIT and DESY (Germany), INFN (Italy), SURF (Netherlands),
PIC (Spain), GridPP (United Kingdom), 
CSCS (Switzerland), IFIN-HH (Romania), CBPF (Brazil),
and Polish WLCG (Poland).
We are indebted to the communities behind the multiple open-source
software packages on which we depend.
Individual groups or members have received support from
ARC and ARDC (Australia);
Key Research Program of Frontier Sciences of CAS, CAS PIFI, CAS CCEPP, 
Fundamental Research Funds for the Central Universities, 
and Sci. \& Tech. Program of Guangzhou (China);
Minciencias (Colombia);
EPLANET, Marie Sk\l{}odowska-Curie Actions, ERC and NextGenerationEU (European Union);
A*MIDEX, ANR, IPhU and Labex P2IO, and R\'{e}gion Auvergne-Rh\^{o}ne-Alpes (France);
AvH Foundation (Germany);
ICSC (Italy); 
GVA, XuntaGal, GENCAT, Inditex, InTalent and Prog.~Atracci\'on Talento, CM (Spain);
SRC (Sweden);
the Leverhulme Trust, the Royal Society
 and UKRI (United Kingdom).

\newpage
\section*{Appendices}
\appendix
\section{Numerical results}\label{app:results}
See Tables~\ref{tab:data_v2_Pt} and~\ref{tab:data_v3_Pt} for the numerical values of the harmonic coefficients $v_2$ and $v_3$.

\begin{table}[ht]
\caption{Numerical values of the harmonic coefficient $v_2(p_T)$. The lower and upper uncertainties are denoted as $\sigma^-$ and $\sigma^+$, respectively.}
\begin{center}
\small{

\label{tab:data_v2_Pt}
\begin{tabular}{c | c c c c c}
\hline\multicolumn{5}{c}{Centrality 65--75\%} \\ \hline
\multirow{2}{*}{mean \pt (\gevc)}& \multirow{2}{*}{$v_2$ ($\times10^2$)}  & \multirow{2}{*}{statistical uncertainty ($\times10^2$)} & \multicolumn{2}{c}{systematic uncertainty}\\ \cline{4-5}
 &  &  & $\sigma^{-}$ ($\times10^2$) & $\sigma^{+}$ ($\times10^2$)\\ \hline
0.3 & 1.293 & 0.027 & 0.27 & 0.30 \\ 
0.5 & 4.328 & 0.027 & 0.23 & 0.30 \\ 
0.7 & 6.885 & 0.031 & 0.29 & 0.34 \\ 
0.9 & 9.060 & 0.038 & 0.36 & 0.32 \\ 
1.2 & 11.708 & 0.035 & 0.094 & 0.15 \\ 
1.7 & 14.232 & 0.056 & 0.083 & 0.21 \\ 
2.2 & 15.375 & 0.090 & 0.16 & 0.18 \\ 
2.7 & 15.22 & 0.14 & 0.19 & 0.18 \\ 
3.4 & 14.33 & 0.17 & 0.16 & 0.16 \\ 
4.4 & 13.12 & 0.34 & 0.47 & 0.47 \\ 
5.7 & 10.3 & 0.5 & 0.4 & 0.6 \\ 
8.0 & 10.3 & 1.0 & 1.2 & 1.1 \\ 
\hline\multicolumn{5}{c}{Centrality 75--84\%} \\ \hline
\multirow{2}{*}{mean \pt (\gevc)}& \multirow{2}{*}{$v_2$ ($\times10^2$)}  & \multirow{2}{*}{statistical uncertainty ($\times10^2$)} & \multicolumn{2}{c}{systematic uncertainty}\\ \cline{4-5}
 &  &  & $\sigma^{-}$ ($\times10^2$) & $\sigma^{+}$ ($\times10^2$)\\ \hline
0.3 & 1.850 & 0.027 & 0.60 & 0.59 \\ 
0.5 & 5.187 & 0.030 & 0.36 & 0.36 \\ 
0.7 & 7.568 & 0.036 & 0.24 & 0.26 \\ 
0.9 & 9.459 & 0.045 & 0.19 & 0.18 \\ 
1.2 & 11.582 & 0.042 & 0.22 & 0.23 \\ 
1.7 & 13.994 & 0.069 & 0.33 & 0.24 \\ 
2.2 & 14.96 & 0.11 & 0.33 & 0.20 \\ 
2.7 & 15.22 & 0.17 & 0.38 & 0.32 \\ 
3.4 & 15.15 & 0.21 & 0.49 & 0.24 \\ 
4.4 & 14.9 & 0.4 & 0.5 & 0.6 \\ 
5.7 & 15.36 & 0.60 & 1.2 & 1.2 \\ 
8.0 & 16.1 & 1.2 & 2.8 & 2.8 \\ 
\hline
\end{tabular}
}
\end{center}
\end{table}
\begin{table}[ht]
\begin{center}
\small{
\caption{Numerical values of the harmonic coefficient $v_3(p_T)$. The lower and upper uncertainties are denoted as $\sigma^-$ and $\sigma^+$, respectively.}
\label{tab:data_v3_Pt}
\begin{tabular}{c | c c c c c}
\hline\multicolumn{5}{c}{Centrality 65--75\%} \\ \hline
\multirow{2}{*}{mean \pt (\gevc)}& \multirow{2}{*}{$v_3$ ($\times10^2$)}  & \multirow{2}{*}{statistical uncertainty ($\times10^2$)} & \multicolumn{2}{c}{systematic uncertainty}\\ \cline{4-5}
 &  &  & $\sigma^{-}$ ($\times10^2$) & $\sigma^{+}$ ($\times10^2$)\\ \hline
0.3 & -0.31 & 0.10 & 0.23 & 0.22 \\ 
0.5 & 1.01 & 0.10 & 0.21 & 0.11 \\ 
0.7 & 2.12 & 0.12 & 0.21 & 0.21 \\ 
0.9 & 2.49 & 0.14 & 0.20 & 0.08 \\ 
1.2 & 3.34 & 0.13 & 0.17 & 0.17 \\ 
1.7 & 4.63 & 0.22 & 0.36 & 0.36 \\ 
2.2 & 5.00 & 0.35 & 0.48 & 0.48 \\ 
2.7 & 4.6 & 0.5 & 0.7 & 0.4 \\ 
3.4 & 2.73 & 0.66 & 0.30 & 0.60 \\ 
4.4 & 2.72 & 1.3 & 0.77 & 0.45 \\ 
5.7 & -1.3 & 1.9 & 2.7 & 2.7 \\ 
8.0 & -5.6 & 3.8 & 1.5 & 1.5 \\ 
\hline\multicolumn{5}{c}{Centrality 75--84\%} \\ \hline
\multirow{2}{*}{mean \pt (\gevc)}& \multirow{2}{*}{$v_3$ ($\times10^2$)}  & \multirow{2}{*}{statistical uncertainty ($\times10^2$)} & \multicolumn{2}{c}{systematic uncertainty}\\ \cline{4-5}
 &  &  & $\sigma^{-}$ ($\times10^2$) & $\sigma^{+}$ ($\times10^2$)\\ \hline
0.3 & -0.90 & 0.14 & 0.26 & 0.30 \\ 
0.5 & 1.13 & 0.15 & 0.28 & 0.28 \\ 
0.7 & 2.14 & 0.19 & 0.24 & 0.12 \\ 
0.9 & 2.20 & 0.23 & 0.28 & 0.28 \\ 
1.2 & 3.32 & 0.22 & 0.16 & 0.15 \\ 
1.7 & 4.15 & 0.36 & 0.24 & 0.20 \\ 
2.2 & 3.10 & 0.56 & 1.3 & 1.3 \\ 
2.7 & 3.0 & 0.9 & 0.6 & 0.6 \\ 
3.4 & 1.67 & 1.1 & 0.74 & 1.2 \\ 
4.4 & -1.0 & 2.1 & 2.2 & 1.6 \\ 
5.7 & -8.7 & 3.0 & 2.6 & 3.2 \\ 
8.0 & -14 & 6 & 8 & 9 \\ 
\hline
\end{tabular}
}
\end{center}
\end{table}
\FloatBarrier



\addcontentsline{toc}{section}{References}
\bibliographystyle{LHCb}
\bibliography{main,standard,LHCb-PAPER,LHCb-CONF,LHCb-DP,LHCb-TDR}

\newpage
\centerline
{\large\bf LHCb collaboration}
\begin
{flushleft}
\small
R.~Aaij$^{35}$\lhcborcid{0000-0003-0533-1952},
A.S.W.~Abdelmotteleb$^{54}$\lhcborcid{0000-0001-7905-0542},
C.~Abellan~Beteta$^{48}$,
F.~Abudin{\'e}n$^{54}$\lhcborcid{0000-0002-6737-3528},
T.~Ackernley$^{58}$\lhcborcid{0000-0002-5951-3498},
B.~Adeva$^{44}$\lhcborcid{0000-0001-9756-3712},
M.~Adinolfi$^{52}$\lhcborcid{0000-0002-1326-1264},
P.~Adlarson$^{78}$\lhcborcid{0000-0001-6280-3851},
C.~Agapopoulou$^{46}$\lhcborcid{0000-0002-2368-0147},
C.A.~Aidala$^{79}$\lhcborcid{0000-0001-9540-4988}
% ...

{\footnotesize \it

% ...
$^{35}$Nikhef National Institute for Subatomic Physics, Amsterdam, Netherlands\\
% ...
$^{44}$Instituto Galego de F{\'\i}sica de Altas Enerx{\'\i}as (IGFAE), Universidade de Santiago de Compostela, Santiago de Compostela, Spain\\
% ...
$^{46}$European Organization for Nuclear Research (CERN), Geneva, Switzerland\\
% ...
$^{48}$Physik-Institut, Universit{\"a}t Z{\"u}rich, Z{\"u}rich, Switzerland\\
% ...
$^{52}$H.H. Wills Physics Laboratory, University of Bristol, Bristol, United Kingdom\\
% ...
$^{54}$Department of Physics, University of Warwick, Coventry, United Kingdom\\
% ...
$^{57}$School of Physics and Astronomy, University of Glasgow, Glasgow, United Kingdom\\
$^{58}$Oliver Lodge Laboratory, University of Liverpool, Liverpool, United Kingdom\\
% ...
$^{66}$Syracuse University, Syracuse, NY, United States\\
% ...
$^{78}$Department of Physics and Astronomy, Uppsala University, Uppsala, Sweden, associated to $^{57}$\\
$^{79}$University of Michigan, Ann Arbor, MI, United States, associated to $^{66}$\\
% ...
}
\end{flushleft}




\end{document}